\documentclass[12pt,preprint]{aastex}

\slugcomment{}

\shorttitle{Starspots in SS Cyg}
\shortauthors{Webb et al.}

\begin{document}


\title{Spectroscopic Evidence for Starspots on the Secondary 
Star of SS Cygni}


\author{N.A. Webb\altaffilmark{1}}
\affil{Centre d'Etude Spatiale des Rayonnements, 9 Avenue du Colonel Roche, 
31028 Toulose Cedex 4, France}
\email{webb@cesr.fr}

\author{T. Naylor\altaffilmark{1}}
\affil{School of Physics, University of Exeter, Stocker Road, Exeter, EX4 4QL,
U.K.}
\email{timn@astro.ex.ac.uk}

\and

\author{R.D. Jeffries}
\affil{Department of Physics, Keele University, Keele, Staffordshire, ST5 5BG,
U.K.}
\email{rdj@astro.keele.ac.uk}


\altaffiltext{1}{Previous address: Department of Physics, Keele University, 
Keele, Staffordshire, ST5 5BG, U.K.}


\begin{abstract}
Cataclysmic variables (CVs) are interacting binary systems where a
cool, rapidly rotating secondary star passes material to a white
dwarf. 
If this mass loss is to continue then there must be continuous
angular momentum loss from the system. 
By analogy with the Sun and other cool stars, it has been assumed that 
magnetic braking is responsible, angular momentum being carried away by 
an ionised wind from the secondary star, threading a dynamo-generated 
magnetic field.
We have discovered TiO absorption in the spectrum of SS Cyg, whose
secondary star should be too hot to show such features.
The most likely explanation of its presence is cool star spots caused by the  
strong (0.1T) fields required by the magnetic braking theories.
\end{abstract}




\section{Introduction}

Cool stars with a spectral type later than F5 (cooler than about
6500K) have convective outer layers, which when coupled with rotation,
create a dynamo that amplifies magnetic fields and brings them to the
stellar surface. Radial magnetic field lines trap the stellar wind, 
forcing it to
co-rotate with the star out to large radii and removing angular
momentum from the star. \citet{01} first applied such a model to the
solar angular momentum loss, and variants of their model can
successfully explain the observed rotational spin-down of cool stars
in open clusters of increasing age \citep{02, 03, 04, 19}.

Low-mass secondary stars in cataclysmic variables (CVs) should exhibit
strong (0.1-0.3T) magnetic fields since they have deep convection zones 
and are
rapidly rotating as a result of being tidally locked with their
companion.  (Tidal interaction ensures that the secondary star is brought
into synchronous rotation with the orbit.)  The currently accepted
hypothesis is that magnetic braking by the secondary star is responsible
for removing angular momentum from the system, perpetuating the mass
transfer \citep{05, 06}. Magnetic braking is also crucial in the
standard explanation for the dearth of CVs with orbital periods
between 2 and 3 hours (the ``period gap'').  As CVs evolve towards
shorter orbital periods the secondary star becomes less massive, until
finally, at masses of 0.25M$_\odot$ and periods around 3 hours, it
should be fully convective \citep{07}.  At this point it is said that either
magnetic activity may cease \citep{08, 06}, or perhaps the magnetic
field changes topology \citep{09}, and thus magnetic braking becomes
ineffectual.  In turn, mass transfer will cease, and the secondary
star, which is slightly bloated due to the effect of mass loss,
shrinks to within its Roche-lobe.  Gravitational radiation is now the
only angular momentum loss mechanism, and only when the binary reaches
a period of two hours does mass transfer begin again.  

There are two worrying problems with this scenario.
First that it relies on applying single star models to semi-detached binaries
at somewhat shorter rotation periods.
The danger of this assumption is underlined by the case of the W UMa binaries,
whose magnetic fields are a factor of order four or five below those for 
analogous single stars \citep{ssv}.
The second problem is the lack of good observational
evidence for magnetic fields in CV secondaries.  
Strong (0.1-0.3T) magnetic fields covering a large fraction of the stellar 
surface are well established in rapidly rotating single and binary G to M-type
stars \citep[e.g.][]{sl}.  
These fields can be detected directly by Zeeman splitting
\citep{10} and inferred indirectly from the presence of cool starspots
on the stellar surface, or from the non-radiative magnetic heating in
outer atmospheres diagnosed by coronal X-ray emission \citep{11, 12,
13}.
Applying these techniques to CVs is difficult, due in part to the variability 
of the accretion disc masking that of the secondary star, and in part
due to their intrinsic faintness.
There are, however, other indications that spots and/or magnetic activity may
be present in CVs.
Magnetic activity cycles are a plausible explanation for changes seen
on timescales of decades in the overall brightness, inter-outburst periods and
orbital periods of CVs.
A critical account of these is given in \cite{rap}, and given that other
explanations are possible, they do not amount to a strong case for magnetic
activity. 
More convincing is to use starspots as an explanation for subtle differences
in secondary star line strengths, either in comparison with single stars of
similar spectral type, or as a function of orbital phase 
\citep[e.g.][]{howell, vik}.
However, there are always other possible explanations for these effects, 
including light from the accretion disc, or the peculiar evolutionary history 
of the secondary star.
In summary, the evidence to date for magnetic activity in CV secondary stars
amounts to using them as possible post-facto explanations of extant 
observational data.
What is required is an observational test of a prediction based on the
idea that the secondary stars have strong magnetic fields.

\section{The technique}

Starspots are the key to the technique we present here.  
Their presence can be deduced from rotational modulation of stellar
brightness or more directly from Doppler imaging of the surfaces of
rapidly rotating stars \citep[e.g.][]{bel}.
\citet{14} have used Zeeman Doppler imaging
to show that starspots do indeed coincide with regions of strong
magnetic field, and starspots are now known to exist in both single
cool stars and non-interacting binaries which have rotation periods
from about 12 hours to a few weeks.  

\citet{11} and \citet{20} have developed a new way to measure the spot
area and temperature, which relies on a typical starspot being
approximately 1000K cooler than the unspotted photosphere. The total
spectrum from the star is then the flux-weighted mean from the spotted
and unspotted areas. In particular, when one observes magnetically
active K stars, molecular absorption features due to TiO are seen,
which can only come from much cooler M-type atmospheres. By modeling
the total spectrum as the sum of template K-type and M-type stars, an
estimate of the spot area and temperature is recovered, assuming that
the star is uniformly covered by spots.  In this letter we present the
first results using a similar method for a CV secondary star.

Although the application of this technique to CVs is straightforward
in principle, there are observational complications which make it
difficult in practice.  First, unlike \citet{11}, we cannot normalise
the spectra such that the continuum flux is unity, as the presence of
non-stellar components (e.g. the accretion disc) whose flux varies with 
wavelength, would make the relative fluxes in the lines incorrect.  
Thus we must flux our
spectra, ensuring that the relative levels across the spectra are
accurate to approximately one percent.  It is well known that normal
techniques can introduce problems at this level, and worse still the
TiO bands are contaminated by absorption from the Earth's
atmosphere. However, preserving this flux information means that our
fitting technique can utilize the overall continuum shape of the
template spectra, in addition to the strengths and widths of the
absorption lines.  For the latter information to be useful, very high
signal-to-noise spectra are required.

\section{Observations and Results}

We chose to investigate the 6.6 hour orbital period CV, SS Cygni. 
Since the secondary star in SS Cyg is an early K star (\citet{h.etal} and
references therein), the mere
detection of M-type TiO features would be a positive indication of
spot activity on its surface.  
Four SS Cyg spectra totaling 24 mins exposure time were taken within 
0.3 in phase of  superior conjunction of the white
dwarf (i.e. when the secondary star is at its closest to the Earth),
using the Intermediate Dispersion Spectrograph on the 2.5m Isaac Newton 
Telescope.
With the 1.6 arcsec slit used, the spectrograph yielded a resolution of 2.5\AA.
The spectra covered the wavelength range 6820-8190\AA\ at a dispersion 
of 1.22\AA\ per pixel.
However, we only used the region 7000-7590\AA, since this was well corrected 
for telluric absorption, and contains the TiO band of interest.
The reduced spectra were shifted so that the spectral lines lay at their 
rest wavelength and added, using a weighted mean, to give a spectrum with a
signal-to-noise of about 200.  
We also observed 35 single K and M
stars to use as template spectra, which were also shifted so that the
spectral lines lay at their rest wavelength positions.  We ensured
good relative flux calibration by observing an O or B-star close to
each target, after the target had been observed.  This hot star was
itself fluxed (using data from flux standards), and the resulting spectrum 
(excluding telluric
absorption bands) was fitted with a quadratic (a good approximation to
the flux distribution of hot stars in this region of the spectrum),
which removed any residual fluxing errors on scales of less than
300\AA.  We next divided the unfluxed spectrum of the target by the
unfluxed spectrum of the O-star, thus removing the telluric absorption
lines.  
Finally we multiplied this spectrum by the quadratic fit to the hot star
spectrum to obtain a fluxed spectrum of the target.

\begin{figure}
\plotone{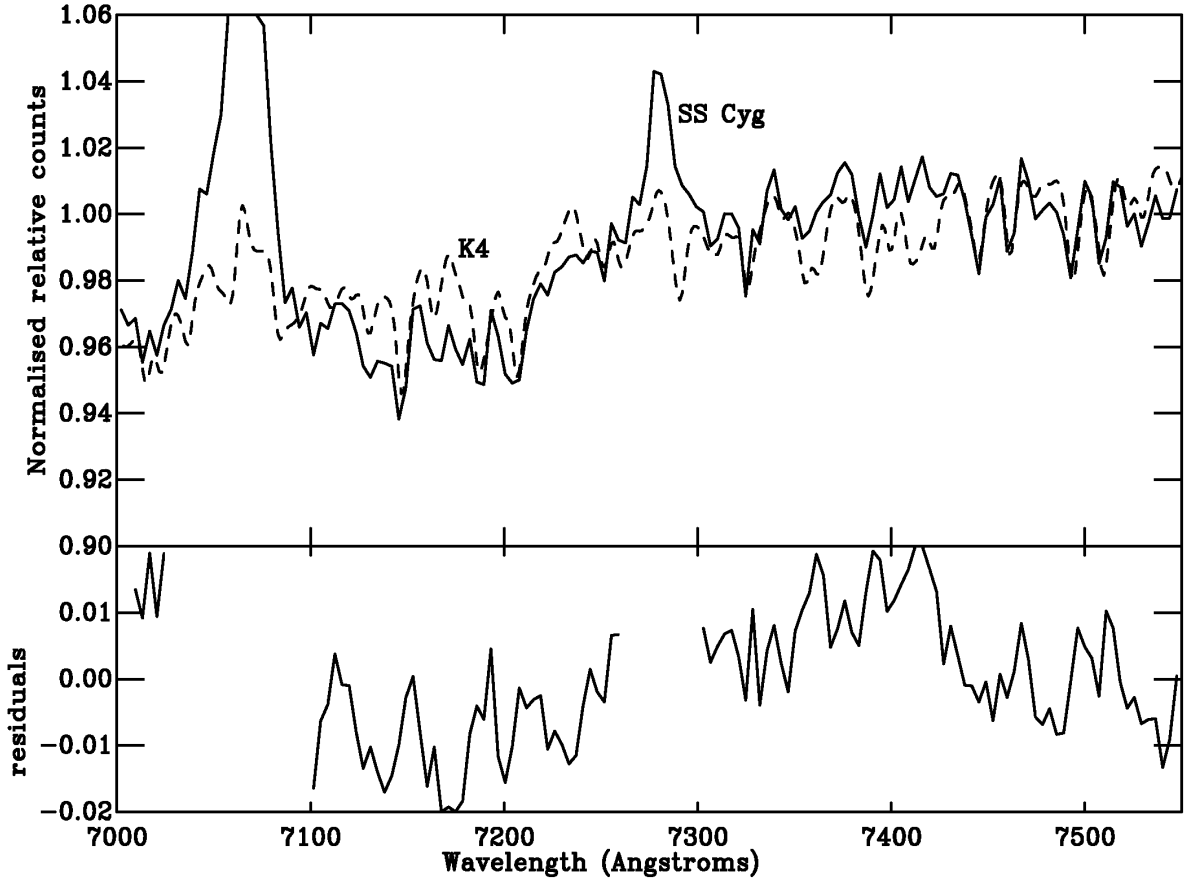}
\caption{
Figure 1.  A K4V star and quadratic only fit (dashed line) to the
Cataclysmic Variable SS Cyg (solid line), between 7000-7590\AA.  
When compared to a simple K4 dwarf spectrum, 
there is a clear depression in the SS Cyg spectrum between 7100\AA\ and 
7200\AA, corresponding to the TiO band, though the bandhead itself (at
7054\AA) is obscured by an emission line.
The bottom panel shows the residuals about the fit.
\label{fig1}}
\end{figure}

Our mean SS Cyg spectrum is shown in Figure \ref{fig1}. 
There are two strong emission lines at 7065\AA\ and 7280\AA\ (both HeI), 
whose broadness implies they originate in the accretion disc.  
(They are broadened both by the rotation of the disc, and our shifting
of the spectra into the rest frame of the secondary star.)
The rest
of the spectrum is a host of weak, rotationally broadened absorption
lines from the secondary star.  
It is immediately evident, even without modeling, that there is a broad 
absorption trough between 7100\AA\ and 7200\AA, which is normally associated 
with TiO  
(although the bandhead itself, at 7055\AA\ is obscured by an emission line).
However, the secondary star in this CV is of spectral type early
K (4800-5300K), and is thus too hot for TiO to exist in significant
quantity.  Indeed TiO only appears in stars of spectral type K7 and
later, where the temperature is below 4300K.  It would appear,
therefore, that we are observing a K-type star, with cooler, M-type
regions, which are presumably starspots.

\begin{figure}
\plotone{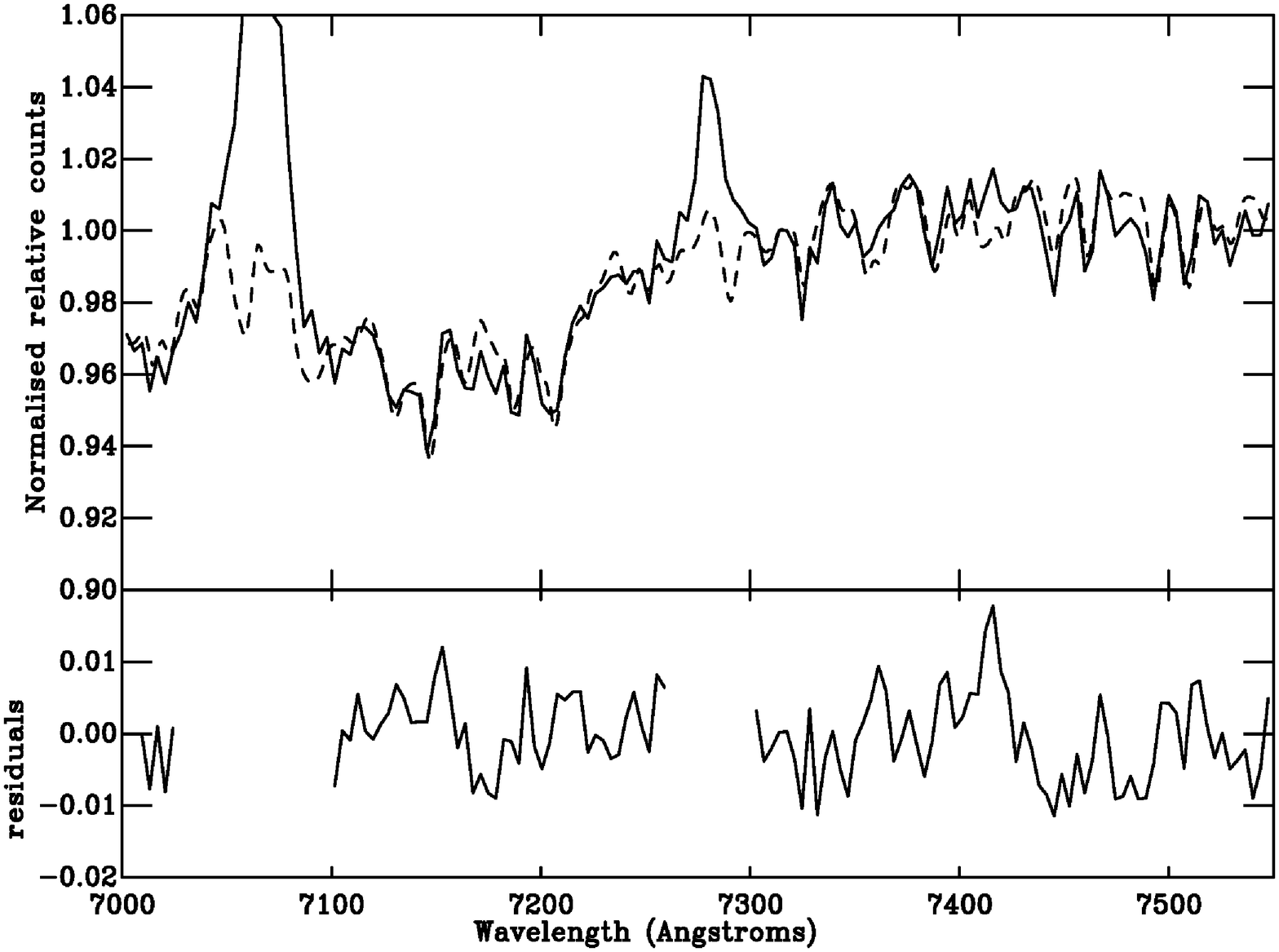}
\caption{
Figure 2.  As figure 1, but the model is now a K4V and M4V star plus
quadratic.  
\label{fig2}}
\end{figure}

We modeled the mean SS Cyg spectrum using varying proportions of our
template K and M, giant and dwarf stars, with the addition of a
shallow quadratic to represent the non-stellar components.  
The spectral lines
of the template stars were broadened to match the lines of SS Cyg
which are rotationally broadened to about 90km s$^{-1}$ \citep{21}.
We restricted the fit to the region 7000-7590\AA, since it contained
one TiO band and was well corrected for telluric
absorption.  To begin with we tried a fit which contained just one
template star, and the quadratic (Figure \ref{fig1}).  
The best fitting spectral type was K4V.
When we added a second star (Figure \ref{fig2}) $\chi^2 _\nu$ fell by
a factor of 2.6.  (There were 126 degrees of freedom for the two star
fit, which returned a $\chi^2 _\nu$ of 3.9.)
The two star model not only fits the area around TiO well, it also fits the
region around 7400\AA\ better than the single star model.
This second fit gave spectral type of K4V
(4800K) for the unspotted parts of the photosphere, and a spot spectral type
M2V (3900K).
The hemisphere we observed had approximately 22\% of its area covered
in spots.
To assess the errors in these figures we fitted a two star model to the 
individual spectra which had been combined to create our mean SS Cyg 
spectrum.
The resulting scatter implied our error is about a 
sub-type for both the ``normal'' and spotted spectral types.
About 50\% of the light was calculated to come from the secondary star.
Surprisingly there was little variation in the implied spot coverage
with changes in either the normal or spotted spectral type.
This is because the fall in flux (as opposed to equivalent width) across
the TiO bandhead changes very little with spectral type, when calculated
per unit area of atmosphere.
The fall in continuum flux being matched
almost exactly by the increasing depth of the absorption.
Using the data of \cite{j.etal} and \cite{r.etal} we find a 10 percent change
in the flux drop across the TiO bandhead between M2 and M4.

To prove the method does not have any propensity to introduce TiO
features where there should not be any, we performed the same
experiment on the known spotted star HD 82443.
In this star the temperature of the unspotted areas is known to be 4925K, 
with a spot temperature of 4300K \citep{22}; thus this object should not 
show detectable TiO.
We fitted the spectrum of this star in an identical fashion to SS Cyg, allowing
both spectral types to run free.
We found no
evidence for TiO, and the fit returned a figure of less than 5\% for
the non-stellar contribution.

\section{Caveats}

The simplest explanation of our observations is that we have observed
starspots on the surface of the secondary star in SS Cyg.
However, there are three other obvious possibilities which should be
considered: gravity darkening
of the secondary star; a peculiar abundance pattern, or an effect caused
by the non-stellar component.

To examine the possibility that gravity darkening may be responsible for the
cooler region(s), we modeled a star with a pole temperature
of 4800K, observed at an inclination of 60 degrees (probable for SS Cyg),
using the code of \cite{snc}.
Gravity darkening will have its greatest effect at phase 0.5, when the inner 
Lagrangian point is facing the observer.
This was our reason for preferring to observe phases close to zero, but to 
ensure our modeling is conservative, we modeled phase 0.3, the closest 
that we observed to 0.5.
We found that only four percent of the flux weighted area lay below the
temperature of 4500K where TiO first becomes visible.
Only if the TiO band depth at this and lower temperatures were 100 percent, 
could our observed depth of 4 percent be explained.
In fact the depth at these temperatures is only a few percent.

Peculiar abundances are also unlikely to cause the observed effect.
The problem is not merely changing the relative strengths of certain lines,
but producing simultaneously the features of an early K star and TiO,
which only occurs in cooler atmospheres.
Modeling a range of abundance possibilities to prove this lies outside the 
scope of this letter, but to be credible, such modeling would have to be 
backed with an explanation as to why the star has the implied abundances.

Finally, could the subtraction of the non-stellar component cause the
TiO features?
Although the exact nature of the non-stellar component is controversial,
it does seem to be smooth on scales of 1000\AA\ and more \citep{h.etal}.
Given that our residuals are also smooth, it seems that only if the
non-stellar component itself has TiO could it explain our results.

\section{Conclusions}

Our conclusion is therefore that the most convincing explanation of 
our data is that starspots do exist on the surfaces of
CV secondary stars. 
These are likely to be associated with equipartition
magnetic fields of strength 0.1-0.3T. This puts the idea that
magnetically controlled wind braking is responsible for CV evolution
on a much firmer footing and opens the way for further studies of
these magnetic fields in ways similar to those being pursued in
rapidly rotating single stars. 
It also shows that magnetic fields and
spot activity do not disappear at extremely high rotation rates, that
are an order of magnitude faster than the claimed dynamo or angular
momentum saturation threshold seen in the K stars of young open
clusters \citep{23}.  
However, much crucial work still needs to be done, including
establishing whether spots disappear at rotation periods of 2-3 hours,
thus providing a natural explanation for the CV period-gap.
CV secondaries at these periods are themselves M-type stars and so the
mere detection of TiO bands will not be sufficient to prove the
presence of starspots.



\acknowledgments

The Isaac Newton Telescope is operated on the island of La Palma by
the Isaac Newton Group in the Spanish Observatorio del Roque de los
Muchachos of the Instituto de Astrofisica de Canarias.  
This work made use of Peter van Hoof's Atomic Line List
(http://www.pa.uky.edu/~peter/atomic).
The majority
of this work was carried out when NAW was supported by a PPARC
studentship at Keele University, and TN was in receipt of a PPARC
Advanced Fellowship at the same institute.
We are grateful to Steve Saar for comments on an earlier version of this
paper.





\begin{thebibliography}{}

\bibitem[Barnes et al.(2001)]
        {bel}Barnes J.R., Collier-Cameron A., James D.J., Donati J.-F., 2001, 
        MNRAS, 324, 231
\bibitem[Bouvier et al.(1997)]
        {03}Bouvier J., Forestini M., Allain S., 1997, \aap, 326, 1023
\bibitem[Dhillon et al.(2002)]
        {vik} Dhillon V.S., Littlefair S.P., Marsh T.R., Sarna M., Boakes E., 2002,
        \apj, accepted
\bibitem[Donati \& Collier Cameron (1997)]
        {14}Donati J-F., Collier Cameron A., 1997, MNRAS, 291, 1
\bibitem[Haisch \& Schmitt (1996)]
        {12}Haisch B.M., Schmitt J.H.M.M., 1996, PASP, 108, 113
\bibitem[Howell et al. (2000)] 
        {howell} Howell S.B., Ciardi D.R., Dhillon V.S., Skidmore, W., 2000,
        \apj 530, 904
\bibitem[Johns-Krull \& Valenti (1996)]
        {10}Johns-Krull C.M., Valenti, J.A., 1996, ApJ, L95
\bibitem[Jones et al (1994)]
        {j.etal} Jones H.R.A., Longmore A.J., Jameson R.F., Mountain C.M.,
        1994, \mnras, 267, 413
\bibitem[Krishnamurthi et al.(1997)]
        {02}Krishnamurthi A., Pinsonneault M.H., Barnes S., Sofia S., 1997, 
        ApJ, 480, 303
\bibitem[Harrison et al. (2000)] 
        {h.etal} Harrison T.E., McNamara B.J., Szkody P., Gilliland R.L.,
        2000, \aj, 120, 2649
\bibitem[Martinez-Pias et al.(1994)]
        {21}Martez-Pias I.G., Giovannelli F., Rossi, C., Gaudenzi S., 1994, 
        \aap, 291, 455
\bibitem[Messina et al.(1999)]
        {22}Messina S., Guinan E.F., Lanza A.F., Ambruster C.,  1999, \aap, 
        347, 249
\bibitem[O'Neal et al.(1996)]
        {11}O'Neal D., Saar S.H., Neff J.E., 1996, ApJ, 463, 766
\bibitem[O'Neal et al.(1998)]
        {20}O'Neal D., Saar S.H., Neff J.E., 1998, \apj, 501, L73
\bibitem[Rappaport et al.(1983)]
        {06}Rappaport S., Joss P.C., Verbunt, F., 1983, ApJ, 275, 713
\bibitem[Reid et al.(1995)]
        {r.etal} Reid I.N., Hawley S.L., Gizis J.E., 1995, \aj, 110, 1838
\bibitem[Richman et al. (1994)]
        {rap} Richman H.R., Applegate J.H., Patterson J., 1994, ApJ, 106, 1075
\bibitem[Robinson et al.(1981)]
        {07}Robinson E.L., Barker E.S., Cochran A.L., Cochran W.D., 
        Nather R.E., 1981, ApJ, 251, 611
\bibitem[Saar \& Linsky (1985)]
        {sl} Saar S.H., Linsky J.L., 1985, ApJ, 299, L47
\bibitem[Shahbaz et al (1993)]
        {snc} Shahbaz T., Naylor T., Charles P.A., 1993, MNRAS, 265, 655
\bibitem[Sills et al.(2000)]
        {19}Sills A., Pinsonneault M.H., Terndrup D.M., 2000, ApJ, 534 335
\bibitem[Spruit \& Ritter (1983)]
        {08}Spruit H.C., Ritter H., 1983, \aap, 124, 267
\bibitem[Stauffer et al.(1994)]
        {23}Stauffer J.R., Caillault J-P., Gange M., Prosser C.F., 
        Hartmann L.W., 1994, ApJSS, 91, 625
\bibitem[St\c{e}pie\'n et al. (2001)]
        {ssv}St\c{e}pie\'n K., Schmidtt J.H.M.M., Voges W., 2001, A\&A, 370, 157
\bibitem[Strassmeier \& Rice (1998)]
        {13}Strassmeier K.G., Rice J.B., 1998, \aap, 330, 685
\bibitem[Taam \& Spruit (1989)]
        {09}Taam R.E., Spruit H.C., 1989, ApJ, 345, 972
\bibitem[Tinker et al.(2000)]
        {04}Tinker J., Pinsonneault M., Terndrup D., 2002, ApJ, 564, 877
\bibitem[Verbunt \& Zwaan (1981)]
        {05}Verbunt F., Zwaan C., 1981, 100, L7
\bibitem[Weber \& Davies (1967)]    
        {01}Weber E., Davis L.D., 1967, ApJ, 148, 217
\end{thebibliography}
\end{document}